\documentclass[twocolumn]{aastex631}
\usepackage{stfloats}
\usepackage{float}
\usepackage{graphicx}
\usepackage{natbib}
\usepackage{hyperref}
\usepackage{amsmath}
\usepackage{autobreak}
\usepackage{multirow}
\usepackage{makecell}
\usepackage{color}
\usepackage{bm}
\usepackage{threeparttable}
\usepackage[figuresright]{rotating}
\usepackage[mathscr]{euscript}
\usepackage[mathlines]{lineno}
\usepackage[utf8]{inputenc}
\usepackage[T1]{fontenc}
\usepackage{amsmath}
\usepackage{makecell}
\usepackage{subfigure}
\usepackage{booktabs}
\usepackage{threeparttable}
\usepackage{needspace}

\begin{document}

\title{A late-time view of the progenitor candidates of the Type~II-P SN\,2009ib and SN\,2012ec}

\correspondingauthor{Ning-Chen Sun; Zexi Niu}
\email{sunnc@ucas.ac.cn; niuzexi@ucas.ac.cn}

\author{Yi-Han Zhao}
\affiliation{School of Astronomy and Space Science, University of Chinese Academy of Sciences, Beijing 100049, China}
\affiliation{National Astronomical Observatories, Chinese Academy of Sciences, Beijing 100101, China}

\author{Xinyi Hong}
\affiliation{School of Astronomy and Space Science, University of Chinese Academy of Sciences, Beijing 100049, China}
\affiliation{National Astronomical Observatories, Chinese Academy of Sciences, Beijing 100101, China}

\author{Ning-Chen Sun}
\affiliation{School of Astronomy and Space Science, University of Chinese Academy of Sciences, Beijing 100049, China}
\affiliation{National Astronomical Observatories, Chinese Academy of Sciences, Beijing 100101, China}
\affiliation{Institute for Frontiers in Astronomy and Astrophysics, Beijing Normal University, Beijing, 102206, China}

\author{Zexi Niu}
\affiliation{School of Astronomy and Space Science, University of Chinese Academy of Sciences, Beijing 100049, China}
\affiliation{National Astronomical Observatories, Chinese Academy of Sciences, Beijing 100101, China}

\author{Justyn R. Maund}
\affiliation{Department of Physics, Royal Holloway, University of London, Egham, TW20 0EX, United Kingdom}

\author{Jifeng Liu}
\affiliation{School of Astronomy and Space Science, University of Chinese Academy of Sciences, Beijing 100049, China}
\affiliation{National Astronomical Observatories, Chinese Academy of Sciences, Beijing 100101, China}
\affiliation{Institute for Frontiers in Astronomy and Astrophysics, Beijing Normal University, Beijing, 102206, China}
\affiliation{New Cornerstone Science Laboratory, National Astronomical Observatories, Chinese Academy of Sciences, Beijing 100012, China}

\begin{abstract}
The progenitors of Type II-P supernovae (SNe) are generally considered to be red supergiants; however, the so-called "red supergiant problem" indicates that a deeper investigation into the progenitors of this class of SNe is necessary. SN\,2009ib and SN\,2012ec are two Type II-P SNe for which progenitor candidates have been identified in pre-explosion images. In this work, we use new, late-time Hubble Space Telescope observations to search for the disappearance of these two candidates and confirm their nature. In the case of SN\,2009ib, the late-time high-resolution imaging reveals that the progenitor candidate is in fact a blend of multiple unresolved stars. Subsequent difference imaging shows no significant change in brightness at the SN’s position even years after the explosion. These findings indicate that the flux from the previously identified source is dominated by unresolved field stars, with little to no contribution from the genuine progenitor. In the case of SN\,2012ec, a comparison of pre-explosion and late-time images reveals that the progenitor candidate faded by about 0.6~mag in the F814W band seven years after the explosion, confirming the disappearance of the progenitor.
\end{abstract}

\section{Introduction} \label{sec:intro}
Core-collapse supernovae (CCSNe) are violent explosive events associated with massive stars undergoing the final stages of their evolution. Most CCSNe are classified as Type~II SNe, characterized by prominent hydrogen lines in their spectra. Type~II-P SNe, a subclass of Type~II SNe, are considered the most common among the CCSNe in the local universe, accounting for $\sim50\%$ of the total (\citealt{smith2011observed}). The designation P refers to the characteristic plateau phase in their post-breakout light curves (\citealt{barbon1979photometric}), which is caused by H recombination (\citealt{grassberg1971theory}).

Type~II-P SNe are widely believed to mainly originate from red supergiant (RSG) progenitors (e.g., \citealt{maund2005progenitor}; \citealt{kilpatrick2018dusty}; \citealt{rui2019probing}; \citealt{van2019type}; \citealt{niu2023dusty}; \citealt{hong2024constraining}), which typically are massive stars with $M > 8\,M_\odot$ and possess massive hydrogen-rich envelopes. This association is strongly supported by direct detections of around 20 Type~II-P SN progenitors in pre-explosion imaging. There is, however, a notable absence of high-mass RSG progenitors ($\sim$17\text{--}25 $M_\odot$) in the observational sample $-$ a discrepancy commonly referred to as the red supergiant problem (\citealt{smartt2009death}). Several explanations have been proposed to account for this discrepancy. One leading hypothesis suggests that stars in this mass range may undergo direct collapse into black holes without producing an observable SN, a process referred to as a failed explosion \citep{o2011black,sukhbold2016core,horiuchi2011cosmic}. Another possibility is that enhanced mass loss $-$ either through stellar winds or eruptive events $-$ removes the hydrogen envelope prior to core collapse, altering the outcome to a stripped-envelope SN or interacting SN rather than a Type~II-P one. Parallel efforts have focused on reassessing the robustness of the problem itself. Interpreting pre-explosion images requires multiple assumptions regarding the progenitor's spectral type, circumstellar extinction, and bolometric correction, all of which introduce significant uncertainties. Additionally, most SN progenitor detections rely on single-band photometry, limiting the accuracy of luminosity estimates, especially for dust-obscured or metal-rich RSGs. Finally, it is important to consider that this discrepancy may not be statistically significant, given the limited number of observed Type II-P progenitors and the low expected frequency of high-mass RSGs according to the initial mass function.


SN\,2009ib is a Type~II-P SN that exploded in the galaxy NGC~1559, for which \citet{takats2015sn} determined a distance of $19.8 \pm 3.0$~Mpc. It was first discovered by the Chilean Automatic Supernova Search (CHASE) project on 2009 August 6.30 UT (\citealt{pignata2009supernova}). \citet{takats2015sn} conducted a detailed analysis of SN\,2009ib. They classified it as a Type~II-P SN, noting its spectra were similar to the subluminous SN\,2002gd, and it exhibited an unusually long plateau phase lasting for about 130\text{--}140~days after explosion. SN\,2009ib showed moderate brightness, with a peak magnitude of $M_V = -15.67 \pm 0.42$~mag, comparable to intermediate-luminosity SNe such as SN\,2008in (\citealt{roy2011sn}) and SN\,2009N (\citealt{takats2014sn}). Although the Galactic reddening in the direction of SN\,2009ib is low ($E(B-V)_{\text{MW}} = 0.0257 \pm 0.0002$~mag, \citealt{schlafly2011measuring}), \citet{takats2015sn} found a host-galaxy reddening of $E(B-V)_{\text{host}} = 0.131 \pm 0.025$~mag, leading to a total reddening of $E(B-V)_{\text{tot}} = 0.16 \pm 0.03$~mag.

In archival pre-explosion images from the Hubble Space Telescope (HST), \citet{takats2015sn} identified a faint source at the position of SN\,2009ib, characterized by a yellow colour ($(V-I)_0 = 0.85$~mag). Assuming this was a single star progenitor, they estimated its initial zero-age main sequence (ZAMS) mass to be $M_{\text{ZAMS}} = 20~M_\odot$. They also considered the possibility that this yellow source was unrelated, and the actual progenitor was a RSG too faint to be detected; in this scenario, they estimated an upper limit for the ZAMS mass of the progenitor to be $\sim 14-17~M_\odot$. 

SN\,2012ec is another Type~II-P SN with an identified progenitor in HST images. First discovered by \citet{monard2012supernova}, it exploded around 2012 August 5 in the spiral galaxy NGC~1084. \citet{barbarino2015sn} adopted a distance modulus of $\mu = 31.19 \pm 0.13$~mag (from \citealt{tully2009extragalactic}) and \citet{maund2013supernova} used a similar Tully-Fisher distance, corresponding to $\sim 17$~Mpc. Analyses of SN\,2012ec were conducted by \citet{maund2013supernova}, \citet{barbarino2015sn}, and \citet{jerkstrand2015supersolar}. Its early spectra showed broad P~Cygni profiles of hydrogen, consistent with a Type~II-P classification, and were similar to SN\,1999em (\citealt{maund2013supernova}). The photometric light curve exhibited a plateau with $M_V \approx -16.54 \pm 0.2$~mag and a duration of $\sim 90$~days (\citealt{barbarino2015sn}); \citet{maund2013supernova} reported $M_{r'} \approx -17.0 \pm 0.1$~mag on the plateau, classifying it as slightly brighter than average for Type~II-P SNe. The Galactic extinction towards SN\,2012ec is $E(B-V))_{\text{MW}} = 0.024$~mag (\citealt{schlafly2011measuring}), while \citet{barbarino2015sn} derived a host-galaxy extinction of $E(B-V))_{\text{host}} = 0.12^{+0.15}_{-0.12}$~mag from Na~I~D lines, resulting in a total reddening of $E(B-V))_{\text{tot}} = 0.144^{+0.15}_{-0.12}$~mag.

In pre-explosion HST F814W images, \citet{maund2013supernova} identified a progenitor candidate. Comparison with MARCS model SEDs, accounting for foreground and host extinction, yielded a progenitor luminosity of $\log(L/L_\odot) = 5.15 \pm 0.19$. From this luminosity, \citet{maund2013supernova} inferred an initial mass range for the progenitor of $14-22~M_\odot$. Separately, nebular phase modelling of [O~I] lines by \citet{jerkstrand2015supersolar} suggested a progenitor $M_{\text{ZAMS}}$ of $13-15~M_\odot$.

The progenitor candidates of SN\,2009ib and SN\,2012ec have been included in many studies of Type~II-P SN progenitors (e.g., \citealp{martinez2020progenitor,goldberg2020value,rodriguez2022luminosity,you2024modeling,van2025red}). Nevertheless, in the absence of late-time imaging, it was difficult to confirm whether these candidates correspond to the genuine progenitors. Now with the new HST late-time images available, it is both timely and necessary to reassess their nature. This work aims to perform such a confirmation and offer direct observational constraints on the progenitor properties for both events. We organize this paper as follows. Section~\ref{sec:data} describes the data used in this work. Sections~\ref{sec:2009ib} and \ref{sec:2012ec} focus on the progenitors of SN\,2009ib and SN\,2012ec, respectively; each section details the methods used to obtain the results, as well as presenting and discussing these results. This paper is then closed with a summary in Section~\ref{sec:sum}.

\section{Data}
\label{sec:data}

The data used in this work are from observations conducted with the HST (see Table~\ref{tab:data} for a complete list). The pre-explosion images of SN\,2009ib were observed with the Wide Field Planetary Camera 2 (WFPC2) using a Wide Field (WF) chip. The late-time images of SN\,2009ib were acquired with the Wide Field Camera 3 (WFC3) Ultraviolet-Visible (UVIS) channel at $t = 6.2\text{--}8.2$~years after the peak brightness. For SN\,2012ec, both the pre-explosion and late-time images were acquired with the Advanced Camera for Surveys (ACS) Wide Field Channel (WFC); the late-time images were taken at $t = 4\text{--}7$~years after the peak brightness. In this work, we used only observations conducted in the F814W filter for easy comparison between different epochs. The HST data were retrieved from the Mikulski Archive for Space Telescopes (MAST; \url{https://mast.stsci.edu}), hosted by the Space Telescope Science Institute.

\begin{table*}[!htbp]
\centering
\caption{HST observations \label{tab:data}}
    \begin{tabular}{cccccccc}
    \hline
    \hline
        \toprule
         Object & Date (UT) & Time From & Instrument & Filter & Exposure & Program & PI\\
          & & Peak (yr) & & & Time (s) & ID & Last Name\\
        \midrule
        SN 2009ib & 2001 Aug 2.5 & $-$8.0 & WFPC2 & F814W & 160 & 9042 & SMARTT\\
         & 2001 Aug 2.5 & $-$8.0 & WFPC2 & F814W & 160 & 9042 & SMARTT\\
         & 2015 Oct 30.6 & 6.2 & WFC3 & F814W & 1365 & 14253 & TAKATS\\
         & 2017 Sep 12.0 & 8.0 & WFC3 & F814W & 930 & 15145 & RIESS\\
         & 2017 Sep 26.0 & 8.1 & WFC3 & F814W & 930 & 15145 & RIESS\\
         & 2017 Nov 7.8 & 8.2 & WFC3 & F814W & 930 & 15145 & RIESS\\
         SN 2012ec & 2010 Sep 20.9 & $-$1.9 & ACS & F814W & 1000 & 11575 & VAN DYK\\
          & 2016 Jul 30.8 & $4.0$ & ACS & F814W & 1728 & 14226 & MAUND\\
          & 2019 Jul 29.1 & $7.0$ & ACS & F814W & 2116 & 15645 & SAND\\

        \bottomrule
    \end{tabular} \\

\end{table*}

\begin{figure*}[htbp]
    \centering
    \includegraphics[width=0.80\textwidth]{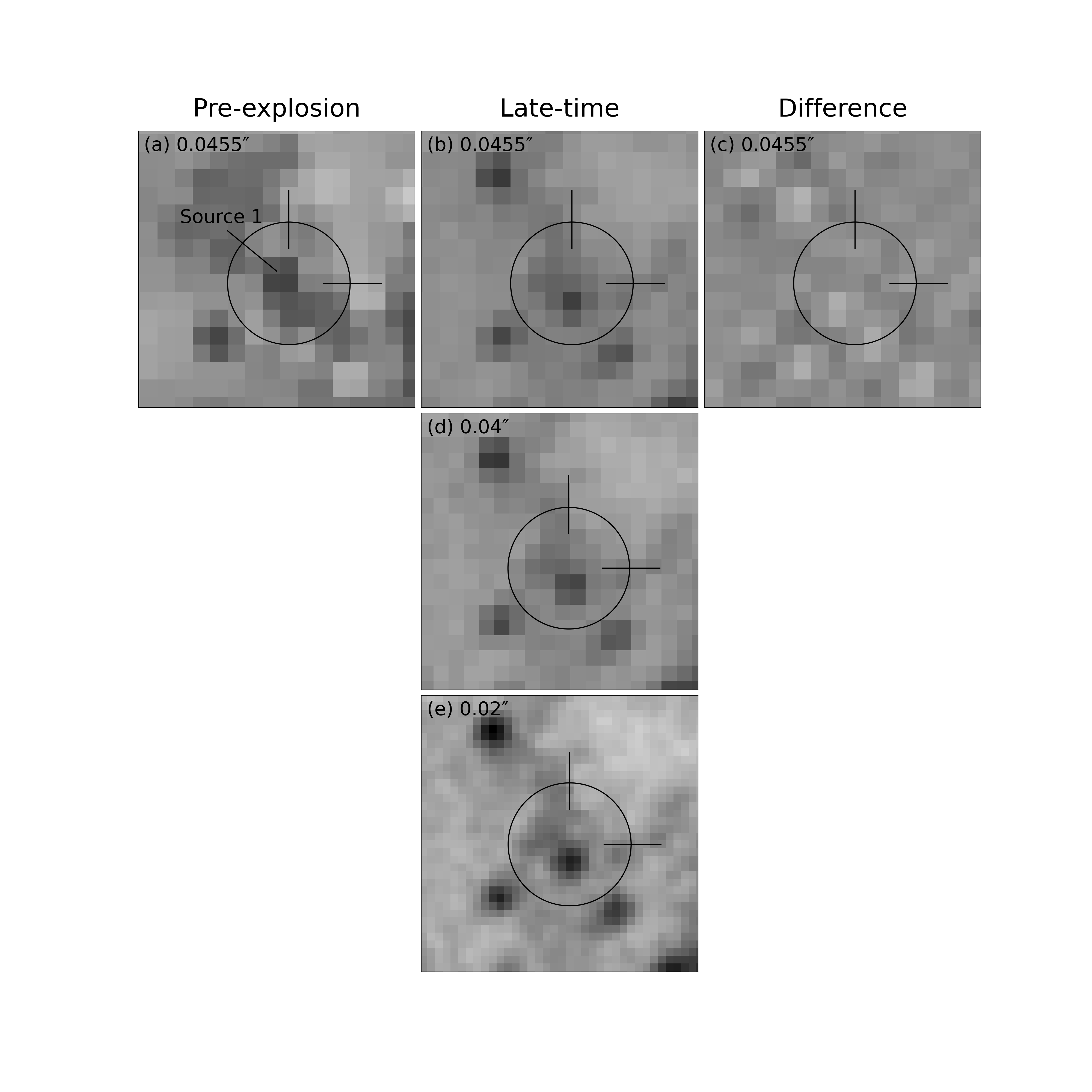}
    \caption{\enspace (a): Pre-explosion HST/WFPC2 image of the site of SN\,2009ib. (b), (d), and (e): Late-time HST/WFC3 images of the SN site with different pixel scales labeled on the images. (c): Difference image obtained by subtracting (b) from (a). All images are in the F814W band. The SN position is marked by a circle in all images with a 1$\sigma$ uncertainty. Each image has a size of $\sim$ 68$\times$68 pc and is oriented with North up and East to the left.}
    \label{fig:2009ib_images}
\end{figure*}

\needspace{2\baselineskip}
\section{SN\,2009\MakeLowercase{ib}}\label{sec:prog}
\label{sec:2009ib}


The pre-explosion image used in this work is from the Hubble Legacy Archive (HLA). The HLA provides uniformly processed, science-ready HST data products based on pipeline-calibrated exposures. These images have been astrometrically aligned, background-matched, and drizzled to correct for geometric distortion. The original instrumental pixel scale is 0.0996", while the HLA image was resampled to an enhanced pixel scale of 0.0455". As shown in the pre-explosion image (Figure~\ref{fig:2009ib_images}a), a source is detected at the position of SN\,2009ib, localized by referencing the images in \citet{takats2015sn}, who identified it as the progenitor candidate of SN\,2009ib. The positional uncertainty, marked by the circle, is 0.3" as noted by \citet{takats2015sn}. For convenience, we refer to this source as \textit{Source~1} throughout this paper.

For the late-time WFC3 images of SN\,2009ib, we first examined the relative alignment between the images and found that they were already well-registered. Then we employed \textsc{astrodrizzle} \citep{hack2012astrodrizzle} package to coadd all late-time images obtained at different epochs. This step also removed cosmic rays and corrected geometric distortion at the same time. The final combined image is shown in Figure~\ref{fig:2009ib_images}d, with a native pixel scale of 0.04". We found there is still significant light at the SN position. The image also indicates that the SN seems to be located in a crowded stellar field.

We further enhanced the image resolution to 0.02"/pixel for a more detailed examination of the SN environment (see Figure~\ref{fig:2009ib_images}e). Here we made full use of the multiple dithered exposures designed with small intentional offsets. The observation conducted in 2015 adopted a 3-point line dither pattern, while each of the three observations conducted in 2017 adopted a 2-point line dither pattern. Due to its imperfect pointing accuracy, the telescope's pointing might exhibit some small offsets across different epochs, leading to an effective 9-point dithering. This enabled the reconstruction of fine spatial details via drizzle-based image combination. In the resulting high-resolution image, \textit{Source~1} is resolved into a dense group of stars.

It should be noted that there is a slight offset between the positions of \textit{Source~1} in the pre-explosion image and the peak brightness within the SN's positional error circle in the late-time image. This could arise from the disappearance of the progenitor star and/or the random brightness fluctuations, and this is further complicated by the different sampling and point-spread functions (PSF) of the images. We resampled the late-time image to a larger pixel scale of 0.045", consistent with the pre-explosion image, for a direct comparison between the two images at matched sampling; then we employed \textsc{swarp} \citep{bertin2010swarp} to align the images at the pixel level, ensuring accurate registration between epochs and minimizing residuals in the subtraction process (see Figure~\ref{fig:2009ib_images}b). We performed image subtraction using the \textsc{hotpants} package \citep{becker2015hotpants}, applied to the pre-explosion and late-time images of SN\,2009ib, which can match the PSFs between the images. This approach allows for a direct examination of any potential brightness change at the SN position.

\begin{figure*}[!ht]
    \centering
    \includegraphics[width=0.80\textwidth]{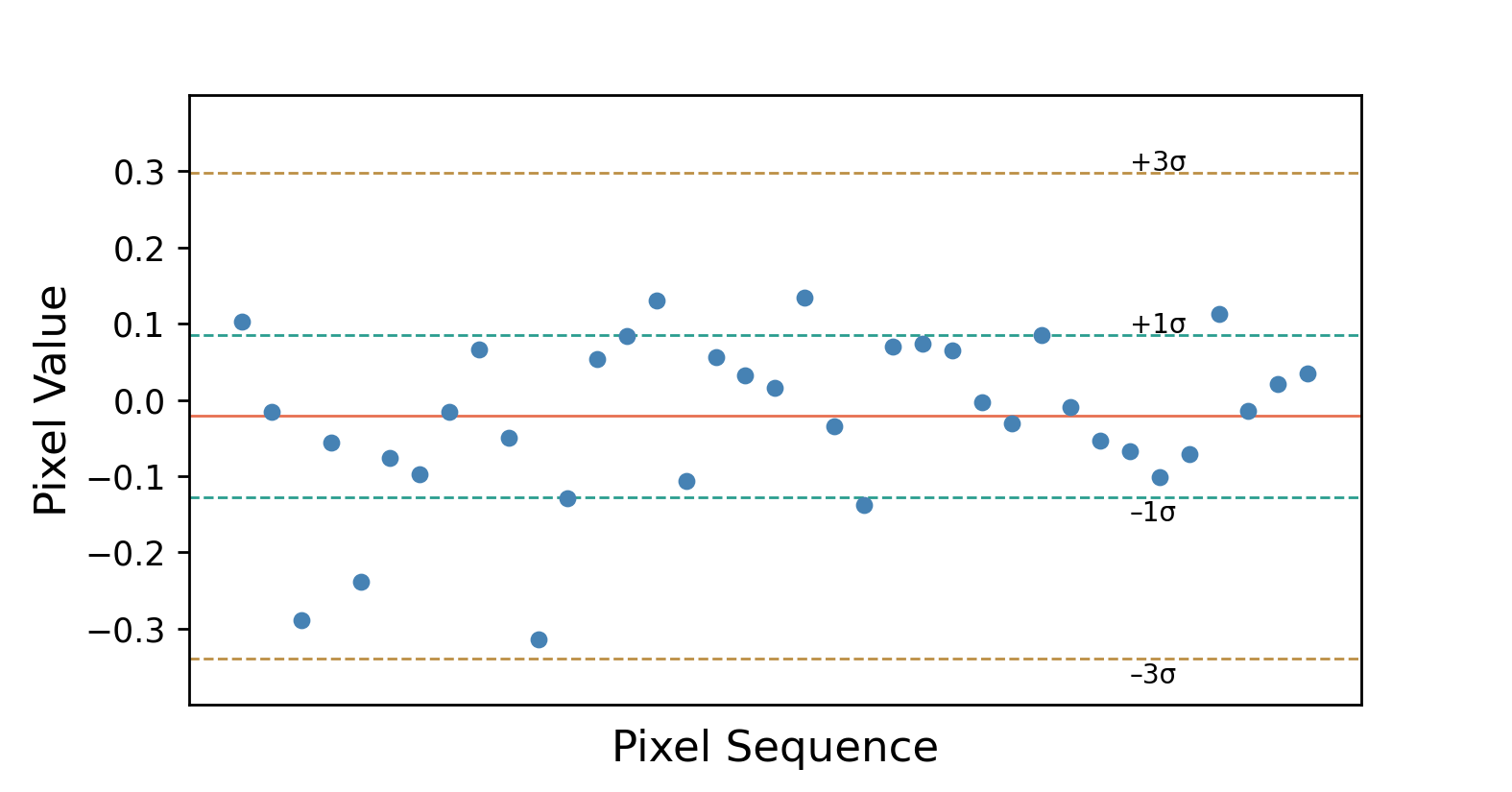}
    \caption{\enspace Values of pixels within the SN's positional error in the difference image shown in Fig.~\ref{fig:2009ib_images}(c). The solid red line marks the zero level, while the dashed lines indicate $\pm 1\sigma$ (teal) and $\pm 3\sigma$ (brown) thresholds. As illustrated in the figure, no signal is detected beyond the $3\sigma$ level.}
    \label{fig:pixel value distribution}
\end{figure*}

\begin{figure*}[!htbp]
    \centering
    \includegraphics[width=1\textwidth]{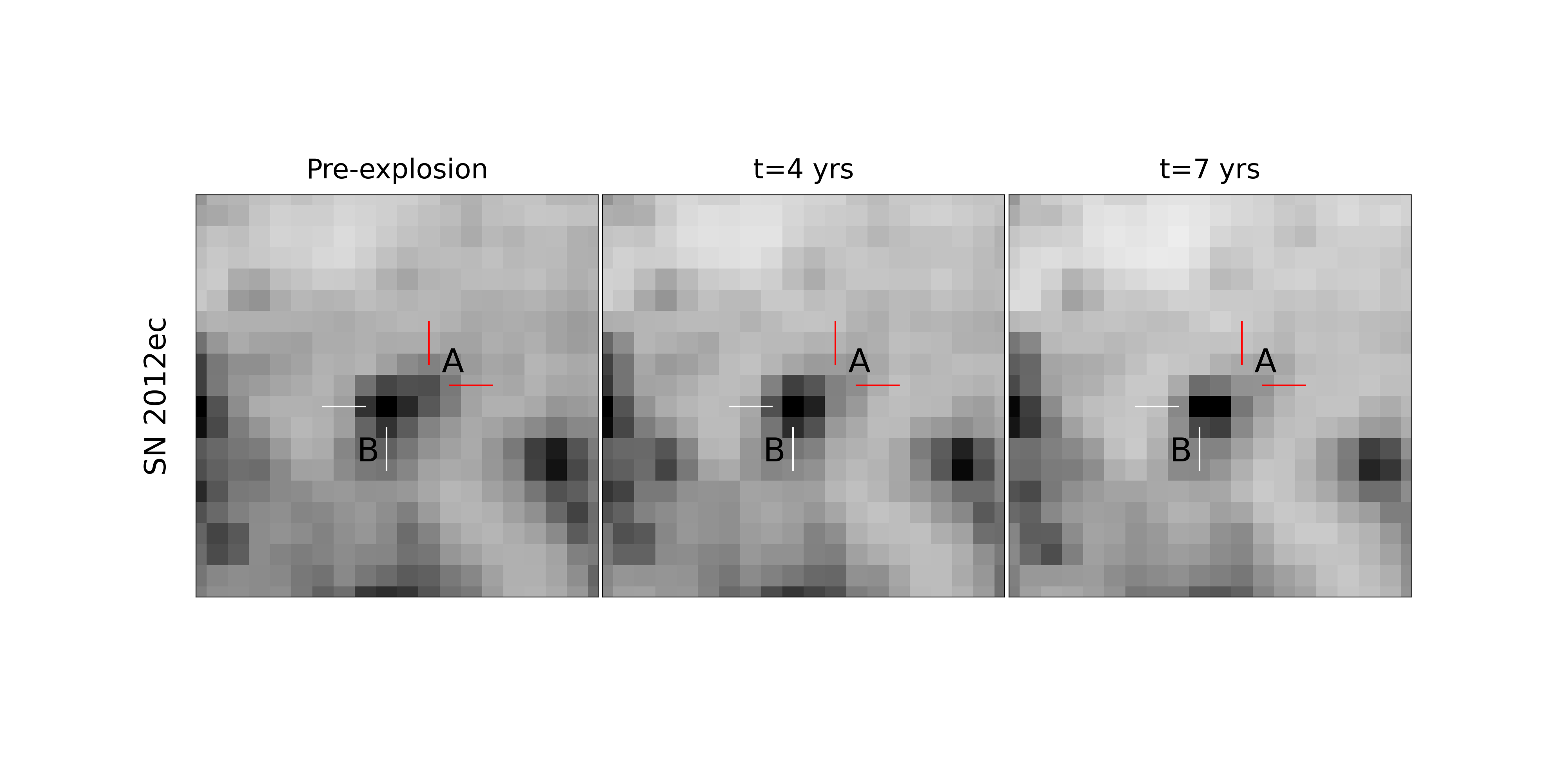}
    \caption{\enspace Pre-explosion and late-time images of the SN\,2012ec site taken with HST/ACS in the F814W filter. The SN position is denoted by the red crosshair, and \textit{Source~B} position by the white crosshair. Each image has a size of $\sim$ 78$\times$78 pc and is oriented with North up and East to the left.}   
    \label{fig:SN2012ec_images}
\end{figure*}

The results of image subtraction are displayed in Figure~\ref{fig:2009ib_images}c and Figure~\ref{fig:pixel value distribution}. No signal beyond the 3$\sigma$ significance level is visible in the difference image, implying no obvious brightness change at the SN position. If \textit{Source~1} were primarily composed of light from the genuine SN progenitor, we would expect a significant decrease in brightness post-explosion, as the progenitor would have vanished. The absence of such a change, however, suggests that \textit{Source~1} is predominantly composed of light from nearby field stars. Furthermore, if the progenitor did contribute to the pre-explosion flux, its contribution must have been minor, insufficient to produce a detectable decrease in brightness after the SN explosion.
Regarding the slight spatial offset in the peak brightness position between the pre-explosion and late-time images, our analysis shows that this offset is not due to an intrinsic change associated with the SN. Instead, it is most likely to originate from random signal fluctuations, possibly noise affecting the centroid of the peak brightness.

\section{SN\,2012\MakeLowercase{ec}}
\label{sec:2012ec}

To process the pre-explosion and late-time epochs of SN\,2012ec, we followed the same alignment and drizzling procedures applied for SN\,2009ib, resulting in cosmic-ray-cleaned, high-quality combined images. Figure~\ref{fig:SN2012ec_images} presents the location of SN\,2012ec as observed in pre-explosion and late-time images. It is clearly shown that there is a source (\textit{Source~A}) at the SN position in the pre-explosion image, with a field source (\textit{Source~B}) adjacent to it. These two sources match those reported in \citet{maund2013supernova}, where \textit{Source~A} was considered as the progenitor candidate of SN\,2012ec. In the late-time images, the light of \textit{Source~A} does not vanish completely, but exhibits a noticeable decrease in brightness. To precisely quantify this change, we conducted point-source photometry on all images using \textsc{dolphot} \citep{dolphin2000wfpc2}. The parameters adopted for photometry were \texttt{FitSky = 2}, \texttt{RAper = 3}, \texttt{ApCor = 0}, \texttt{Force1 = 1}, as recommended for crowded fields. We derive a pre-explosion magnitude of $m_{\mathrm{F814W}}$ = 23.20 $\pm$ 0.04~mag for \textit{Source~A}. This is consistent with 23.10 $\pm$ 0.04~mag measured by \citet{maund2013supernova} from ACS/WFC imaging, within almost 1$\sigma$ uncertainties. They also reported a fainter value of $m_{\mathrm{F814W}} = 23.39 \pm 0.18$~mag from pre-explosion HST/WFPC2 data, noting that its lower precision could be due to difficulties in deblending \textit{Source~A} from \textit{Source~B} and also the slight difference
between the F814W filters used by WFPC2 and ACS. The late-time magnitudes at $t = 4$ and 7~yrs for \textit{Source~A} are derived to be $m_{\mathrm{F814W}}$ = 23.70 $\pm$ 0.06 and $m_{\mathrm{F814W}}$ = 23.84 $\pm$ 0.06, respectively. The magnitude derived for \textit{Source~B} is $m_{\mathrm{F814W}}$ = 22.28 $\pm$ 0.01~mag, also consistent with the $m_{\mathrm{F814W}}$ = 22.32 $\pm$ 0.02~mag in \citep{maund2013supernova} within the uncertainties.

\begin{figure*}[!htbp]
    \centering
    \includegraphics[width=0.80\textwidth]{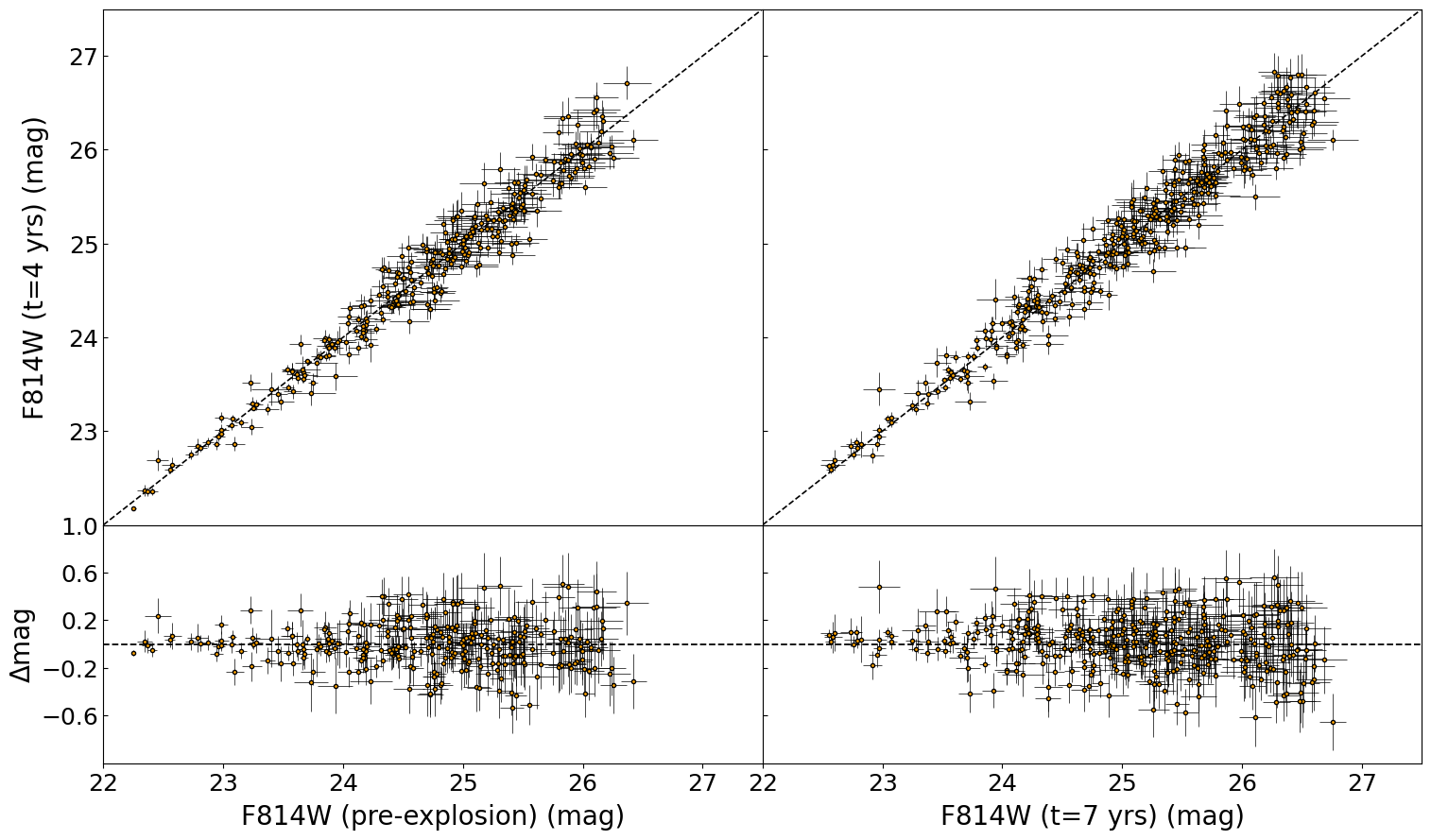}
    \caption{\enspace Comparison of photometry as reported by \textsc{dolphot} for observations at different epochs. The black dashed line shows the one-to-one relation. No significant systematic shifts are evident.}
    \label{fig:syserr}
\end{figure*}

\begin{figure*}[!htbp]
    \centering
    \includegraphics[width=0.80\textwidth]{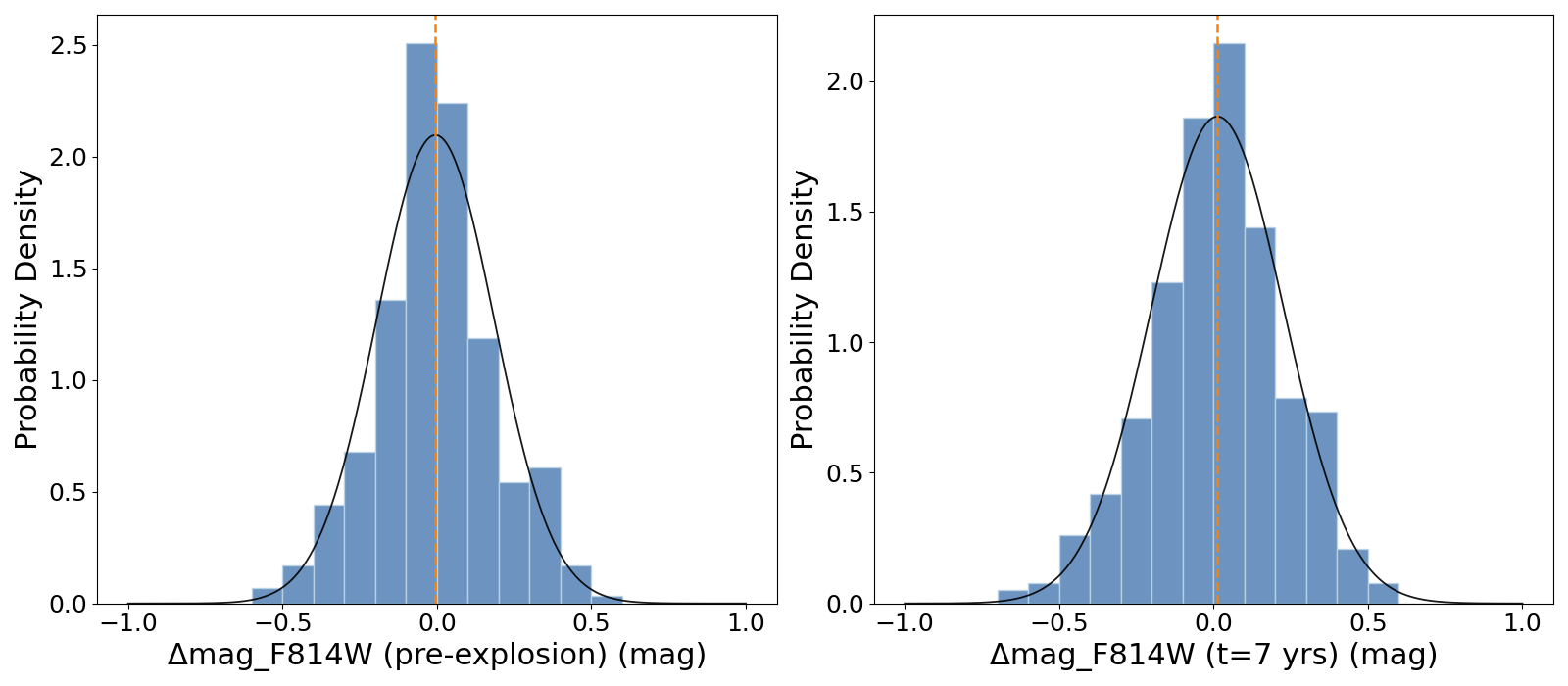}
    \caption{\enspace Histograms of magnitude differences with respect to the epoch of t = 4 yrs. The solid lines are Gaussian fit curves and the dashed lines represent the means of the Gaussian fit. Normalization has been applied to each graph.}
    \label{fig:histogram}
\end{figure*}

To correct for possible systematic biases between different epochs, we selected common stars and compared their photometric measurements. The observations taken at the second epoch ($t = 4$ years) served as the reference frame. With a signal-to-noise ratio threshold of 5, we selected several hundred stars for comparison. We calculated inverse-variance weighted averages of the magnitude differences between different epochs, applying iterative 3$\sigma$ clipping to minimize outlier influence and obtain robust estimates of systematic errors. The results, shown in Figure~\ref{fig:syserr} and Figure~\ref{fig:histogram}, indicate no significant systematic deviations among the three epochs. This confirms the overall consistency of the photometric data.

\begin{figure*}[!htbp]
    \centering
    \includegraphics[width=0.80\textwidth]{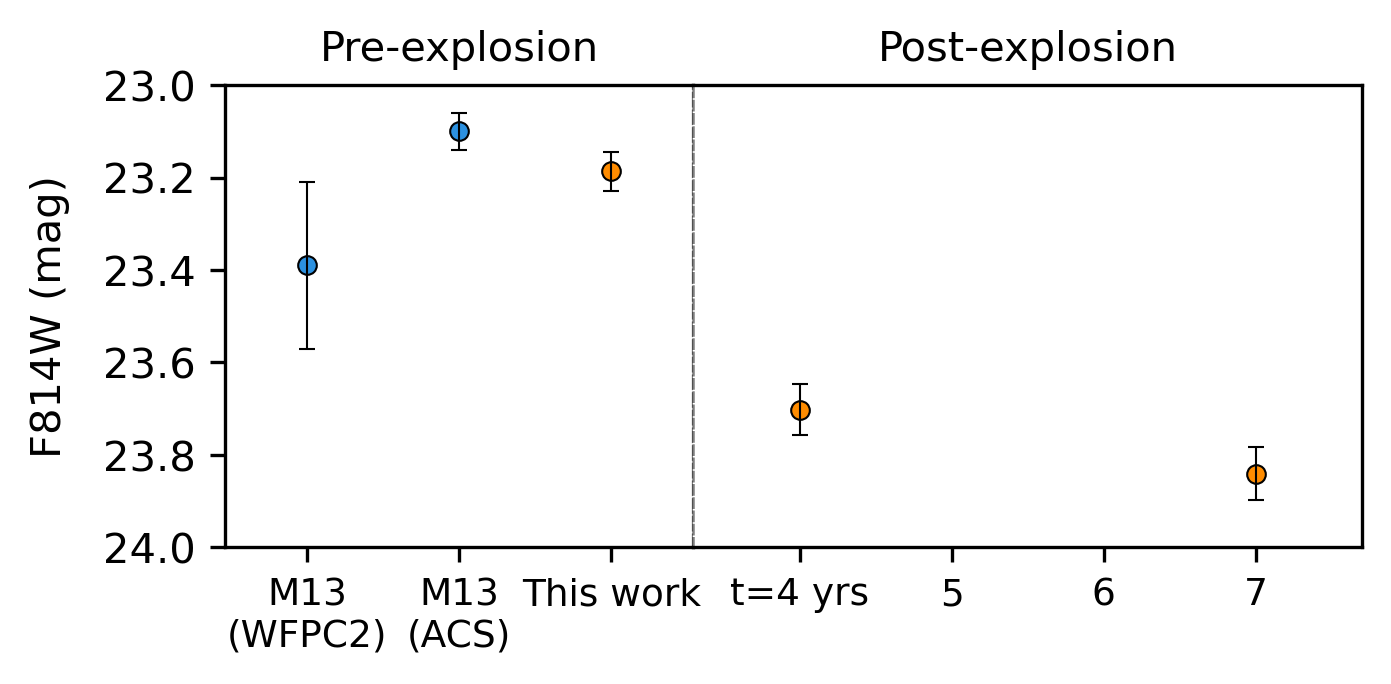}
    \caption{\enspace Light curve of \textit{Source~A}. The pre-explosion magnitudes measured with WFPC2 and ACS by \citet{maund2013supernova} are also plotted for comparison, labelled with M13 (WFPC2) and M13 (ACS) on the X-axis, respectively.}
    \label{fig:light curves-SN2012ec}
\end{figure*}

We then constructed the light curve of \textit{Source~A}, as shown in Figure~\ref{fig:light curves-SN2012ec}. The brightness of \textit{Source~A} declined by approximately 0.6~mag seven years after the SN explosion, and shows evidence of an ongoing, gradual fading. According to the above results, we suggest that \textit{Source~A} is the genuine progenitor of SN\,2012ec and has disappeared after the SN explosion. The residual light detected at the SN location in the late-time images may originate from different mechanisms. One possibility is a light echo, where SN light is scattered by interstellar or circumstellar dust into the line of sight. Such echoes can persist for years after the explosion, producing a slowly fading, spatially extended signal that may contaminate photometric measurements of the progenitor site (\citealt{Boffi1999}; \citealt{bond2003energetic}). Another possible explanation is continued interaction between the SN ejecta and the surrounding circumstellar material (CSM). This interaction can give rise to late-time emission, especially in radio, optical and X-ray wavelengths, as the fast-moving ejecta shock and heat the CSM (\citealt{Fraser2020}; \citealt{Pellegrino_2022}). The resulting emission typically fades over time as the available CSM is exhausted or becomes optically thin. 

\section{Summary}
\label{sec:sum}

In this work, we revisited the progenitor candidates previously identified for SN\,2009ib and SN\,2012ec. For SN\,2009ib, our analysis shows that the reported candidate is not the genuine progenitor, but rather dominated by light from a dense group of unrelated field stars, which were unresolved in the pre-explosion image. No significant brightness change is detected at the SN position even years after the explosion, further supporting the conclusion that the actual progenitor has not been directly identified. In contrast, for SN\,2012ec, we confirm that the previously proposed source is indeed the progenitor, and has disappeared following the SN explosion. The residual emission at the SN site may originate from post-explosion processes, such as a light echo produced by dust scattering or continued interaction between the SN ejecta and surrounding CSM. Our results reinforce the importance of high-resolution imaging and late-time follow-up in confirming the identity and fate of SN progenitors.

\section*{acknowledgments}
NCS’s research is funded by the NSFC grants No. 12303051 and No. 12261141690 and ZXN acknowledges support from the NSFC through grant No. 12303039. JFL acknowledges support from the NSFC through grants No. 11988101 and No. 11933004 and from the New Cornerstone Science Foundation through the New Cornerstone Investigator Program and the XPLORER PRIZE. This work is supported by the Strategic Priority Research Program of the Chinese Academy of Sciences, Grant No. XDB0550300, and by the China Manned Space Program with Grant No. CMS-CSST-2025-A14. The specific observations analyzed in this work are available via \href{https://doi.org/10.17909/42er-4v78}{MAST} at DOI: \href{https://doi.org/10.17909/42er-4v78}{10.17909/42er-4v78}.

\clearpage

\vspace{5mm}

\bibliography{SN2012ec}

\begin{thebibliography}{}
\expandafter\ifx\csname natexlab\endcsname\relax\def\natexlab#1{#1}\fi
\providecommand{\url}[1]{\href{#1}{#1}}
\providecommand{\dodoi}[1]{doi:~\href{http://doi.org/#1}{\nolinkurl{#1}}}
\providecommand{\doeprint}[1]{\href{http://ascl.net/#1}{\nolinkurl{http://ascl.net/#1}}}
\providecommand{\doarXiv}[1]{\href{https://arxiv.org/abs/#1}{\nolinkurl{https://arxiv.org/abs/#1}}}

\bibitem[{Barbarino {et~al.}(2015)Barbarino, Dall'Ora, Botticella, Valle, Zampieri, Maund, Pumo, Jerkstrand, Benetti, Elias-Rosa, {et~al.}}]{barbarino2015sn}
Barbarino, C., Dall'Ora, M., Botticella, M., {et~al.} 2015, Monthly Notices of the Royal Astronomical Society, 448, 2312

\bibitem[{Barbon {et~al.}(1979)Barbon, Ciatti, \& Rosino}]{barbon1979photometric}
Barbon, R., Ciatti, F., \& Rosino, L. 1979, Astronomy and Astrophysics, vol. 72, no. 3, Feb. 1979, p. 287-292. Research supported by the Consiglio Nazionale delle Ricerche., 72, 287

\bibitem[{Becker(2015)}]{becker2015hotpants}
Becker, A. 2015, Astrophysics Source Code Library, ascl

\bibitem[{Bertin(2010)}]{bertin2010swarp}
Bertin, E. 2010, Astrophysics Source Code Library, ascl

\bibitem[{Boffi {et~al.}(1999)Boffi, Sparks, \& Macchetto}]{Boffi1999}
Boffi, F.~R., Sparks, W.~B., \& Macchetto, F.~D. 1999, Astronomy and Astrophysics Supplement Series, 138, 253

\bibitem[{Bond {et~al.}(2003)Bond, Henden, Levay, Panagia, Sparks, Starrfield, Wagner, Corradi, \& Munari}]{bond2003energetic}
Bond, H.~E., Henden, A., Levay, Z.~G., {et~al.} 2003, Nature, 422, 405

\bibitem[{Dolphin(2000)}]{dolphin2000wfpc2}
Dolphin, A.~E. 2000, Publications of the Astronomical Society of the Pacific, 112, 1383

\bibitem[{Fraser(2020)}]{Fraser2020}
Fraser, M. 2020, Royal Society Open Science, 7, 200467

\bibitem[{Goldberg \& Bildsten(2020)}]{goldberg2020value}
Goldberg, J.~A., \& Bildsten, L. 2020, The Astrophysical Journal Letters, 895, L45

\bibitem[{Grassberg {et~al.}(1971)Grassberg, Imshennik, \& Nadyozhin}]{grassberg1971theory}
Grassberg, E., Imshennik, V., \& Nadyozhin, D. 1971, Astrophysics and Space Science, 10, 28

\bibitem[{Hack {et~al.}(2012)Hack, Dencheva, Fruchter, Armstrong, Avila, Baggett, Bray, Droettboom, Dulude, Gonzaga, {et~al.}}]{hack2012astrodrizzle}
Hack, W.~J., Dencheva, N., Fruchter, A., {et~al.} 2012, in American Astronomical Society Meeting Abstracts\# 220, Vol. 220, 135--15

\bibitem[{Hong {et~al.}(2024)Hong, Sun, Niu, Wu, Xi, \& Liu}]{hong2024constraining}
Hong, X., Sun, N.-C., Niu, Z., {et~al.} 2024, The Astrophysical Journal Letters, 977, L50

\bibitem[{Horiuchi {et~al.}(2011)Horiuchi, Beacom, Kochanek, Prieto, Stanek, \& Thompson}]{horiuchi2011cosmic}
Horiuchi, S., Beacom, J.~F., Kochanek, C.~S., {et~al.} 2011, The Astrophysical Journal, 738, 154

\bibitem[{Jerkstrand {et~al.}(2015)Jerkstrand, Smartt, Sollerman, Inserra, Fraser, Spyromilio, Fransson, Chen, Barbarino, Dall'Ora, {et~al.}}]{jerkstrand2015supersolar}
Jerkstrand, A., Smartt, S.~J., Sollerman, J., {et~al.} 2015, Monthly Notices of the Royal Astronomical Society, 448, 2482

\bibitem[{Kilpatrick \& Foley(2018)}]{kilpatrick2018dusty}
Kilpatrick, C.~D., \& Foley, R.~J. 2018, Monthly Notices of the Royal Astronomical Society, 481, 2536

\bibitem[{Martinez {et~al.}(2020)Martinez, Bersten, Anderson, Gonz{\'a}lez-Gait{\'a}n, F{\"o}rster, \& Folatelli}]{martinez2020progenitor}
Martinez, L., Bersten, M.~C., Anderson, J.~P., {et~al.} 2020, Astronomy \& Astrophysics, 642, A143

\bibitem[{Maund {et~al.}(2013)Maund, Fraser, Smartt, Botticella, Barbarino, Childress, Gal-Yam, Inserra, Pignata, Reichart, {et~al.}}]{maund2013supernova}
Maund, J., Fraser, M., Smartt, S.~J., {et~al.} 2013, Monthly Notices of the Royal Astronomical Society: Letters, 431, L102

\bibitem[{Maund {et~al.}(2005)Maund, Smartt, \& Danziger}]{maund2005progenitor}
Maund, J.~R., Smartt, S.~J., \& Danziger, I.~J. 2005, Monthly Notices of the Royal Astronomical Society: Letters, 364, L33

\bibitem[{Monard {et~al.}(2012)Monard, Childress, Scalzo, Yuan, \& Schmidt}]{monard2012supernova}
Monard, L., Childress, M., Scalzo, R., Yuan, F., \& Schmidt, B. 2012, Central Bureau Electronic Telegrams, 3201, 1

\bibitem[{Niu {et~al.}(2023)Niu, Sun, Maund, Zhang, Zhao, \& Liu}]{niu2023dusty}
Niu, Z., Sun, N.-C., Maund, J.~R., {et~al.} 2023, The Astrophysical Journal Letters, 955, L15

\bibitem[{O'Connor \& Ott(2011)}]{o2011black}
O'Connor, E., \& Ott, C.~D. 2011, The Astrophysical Journal, 730, 70

\bibitem[{Pellegrino {et~al.}(2022)Pellegrino, Howell, Vinkó, Gangopadhyay, Xiang, Arcavi, Brown, Burke, Hiramatsu, Hosseinzadeh, Li, McCully, Misra, Newsome, Gonzalez, Pritchard, Valenti, Wang, \& Zhang}]{Pellegrino_2022}
Pellegrino, C., Howell, D.~A., Vinkó, J., {et~al.} 2022, The Astrophysical Journal, 926, 125, \dodoi{10.3847/1538-4357/ac3e63}

\bibitem[{Pignata {et~al.}(2009)Pignata, Maza, Hamuy, Antezana, Gonzalez, Gonzalez, Lopez, Silva, Folatelli, Iturra, {et~al.}}]{pignata2009supernova}
Pignata, G., Maza, J., Hamuy, M., {et~al.} 2009, Central Bureau Electronic Telegrams, 1902, 1

\bibitem[{Rodr{\'\i}guez(2022)}]{rodriguez2022luminosity}
Rodr{\'\i}guez, {\'O}. 2022, Monthly Notices of the Royal Astronomical Society, 515, 897

\bibitem[{Roy {et~al.}(2011)Roy, Kumar, Benetti, Pastorello, Yuan, Brown, Immler, Fatkhullin, Moskvitin, Maund, {et~al.}}]{roy2011sn}
Roy, R., Kumar, B., Benetti, S., {et~al.} 2011, The Astrophysical Journal, 736, 76

\bibitem[{Rui {et~al.}(2019)Rui, Wang, Mo, Xiang, Zhang, Maund, Gal-Yam, Wang, \& Zhang}]{rui2019probing}
Rui, L., Wang, X., Mo, J., {et~al.} 2019, Monthly Notices of the Royal Astronomical Society, 485, 1990

\bibitem[{Schlafly \& Finkbeiner(2011)}]{schlafly2011measuring}
Schlafly, E.~F., \& Finkbeiner, D.~P. 2011, The Astrophysical Journal, 737, 103

\bibitem[{Smartt {et~al.}(2009)Smartt, Eldridge, Crockett, \& Maund}]{smartt2009death}
Smartt, S., Eldridge, J., Crockett, R., \& Maund, J. 2009, Monthly Notices of the Royal Astronomical Society, 395, 1409

\bibitem[{Smith {et~al.}(2011)Smith, Li, Filippenko, \& Chornock}]{smith2011observed}
Smith, N., Li, W., Filippenko, A.~V., \& Chornock, R. 2011, Monthly Notices of the Royal Astronomical Society, 412, 1522

\bibitem[{Sukhbold {et~al.}(2016)Sukhbold, Ertl, Woosley, Brown, \& Janka}]{sukhbold2016core}
Sukhbold, T., Ertl, T., Woosley, S., Brown, J.~M., \& Janka, H.-T. 2016, The Astrophysical Journal, 821, 38

\bibitem[{Tak{\'a}ts {et~al.}(2014)Tak{\'a}ts, Pumo, Elias-Rosa, Pastorello, Pignata, Paillas, Zampieri, Anderson, Vink{\'o}, Benetti, {et~al.}}]{takats2014sn}
Tak{\'a}ts, K., Pumo, M., Elias-Rosa, N., {et~al.} 2014, Monthly Notices of the Royal Astronomical Society, 438, 368

\bibitem[{Tak{\'a}ts {et~al.}(2015)Tak{\'a}ts, Pignata, Pumo, Paillas, Zampieri, Elias-Rosa, Benetti, Bufano, Cappellaro, Ergon, {et~al.}}]{takats2015sn}
Tak{\'a}ts, K., Pignata, G., Pumo, M., {et~al.} 2015, Monthly Notices of the Royal Astronomical Society, 450, 3137

\bibitem[{Tully {et~al.}(2009)Tully, Rizzi, Shaya, Courtois, Makarov, \& Jacobs}]{tully2009extragalactic}
Tully, R.~B., Rizzi, L., Shaya, E.~J., {et~al.} 2009, The Astronomical Journal, 138, 323

\bibitem[{Van~Dyk(2025)}]{van2025red}
Van~Dyk, S.~D. 2025, Galaxies (2075-4434), 13

\bibitem[{Van~Dyk {et~al.}(2019)Van~Dyk, Zheng, Maund, Brink, Srinivasan, Andrews, Smith, Leonard, Morozova, Filippenko, {et~al.}}]{van2019type}
Van~Dyk, S.~D., Zheng, W., Maund, J.~R., {et~al.} 2019, The Astrophysical Journal, 875, 136

\bibitem[{You {et~al.}(2024)You, Chen, Pan, Tsai, \& Ou}]{you2024modeling}
You, K.-A., Chen, K.-J., Pan, Y.-C., Tsai, S.-H., \& Ou, P.-S. 2024, The Astrophysical Journal, 970, 145

\end{thebibliography}
\bibliographystyle{aasjournal}

\end{document}